\documentclass[twoside]{article}
\usepackage{qic,epsfig}
\usepackage{braket}
\usepackage{txfonts}
\usepackage{xcolor}
\newtheorem{proposition}{Proposition}
\newtheorem{remark}{Remark}
\newtheorem{question}{Question}

\begin{document}
\setlength{\textheight}{8.0truein}    

\runninghead{Unextendibility, uncompletability, and many-copy indistinguishable ensembles}{Saronath Halder and Alexander Streltsov}

\normalsize\textlineskip
\thispagestyle{empty}
\setcounter{page}{1}
\alphfootnote
\fpage{1}

\centerline{\bf UNEXTENDIBILITY, UNCOMPLETABILITY, AND MANY-COPY}
\vspace*{0.035truein}
\centerline{\bf INDISTINGUISHABLE ENSEMBLES}
\vspace*{0.37truein}

\centerline{\footnotesize SARONATH HALDER$^1$, ALEXANDER STRELTSOV$^2$}
\vspace*{0.015truein}

\centerline{\footnotesize\it $^1$Centre for Quantum Optical Technologies, Centre of New Technologies, University of Warsaw}
\baselineskip=10pt
\centerline{\footnotesize\it Banacha 2c, 02-097 Warsaw, Poland} 

\centerline{\footnotesize\it $^2$Institute of Fundamental Technological Research, Polish Academy of Sciences}
\baselineskip=10pt
\centerline{\footnotesize\it Pawi\'{n}skiego 5B, 02-106 Warsaw, Poland}
\vspace*{0.21truein}

\abstracts{In this work, we explore the notions unextendible product basis and uncompletability for operators which remain positive under partial transpose. Then, we analyze their connections to the ensembles which are many-copy indistinguishable under local operations and classical communication (LOCC). We show that the orthogonal complement of any bipartite pure entangled state is spanned by product states which form a nonorthogonal unextendible product basis (nUPB) of maximum cardinality. This subspace has one to one correspondence with the maximum dimensional subspace where there is no orthonormal product basis. Due to these, the proof of indistinguishability of a class of ensembles under LOCC in many-copy scenario becomes simpler. Furthermore, it is now clear that there are several many-copy indistinguishable ensembles which are different construction-wise. But if we consider the technique of proving their indistinguishability property under LOCC, then, for many of them it can be done using the general notion of unextendible product basis. Explicit construction of the product states, forming nUPBs is shown. Thereafter, we introduce the notion of positive partial transpose uncompletability to unify different many-copy indistinguishable ensembles. We also report a class of multipartite many-copy indistinguishable ensembles for which local indistinguishability property increases with decreasing number of mixed states.}{}{}

\vspace*{10pt}
\vspace*{3pt}\textlineskip

\section{Introduction}\label{sec1}
A composite quantum system may exhibit nonlocal properties. Such properties are of central importance in quantum information theory since they make quantum systems fundamentally different from their classical counterparts. Over the years, researchers have reported many nonlocal properties. One such property is associated with the state discrimination problem under local operations and classical communication (LOCC). In brief, it is called the local state discrimination problem (LSDP). In this problem, a composite quantum system is prepared in a state which is secretly taken from a given set. The task is to identify that state by LOCC. 

For orthogonal states, it is always possible to identify the state of the system via a suitable global measurement. But it is not always possible to identify the state of the system perfectly under the setting of LSDP even if the given set only contains orthogonal states \cite{Bennett99-1, Bennett99, Walgate00, Virmani01, Ghosh01, Walgate02, Ghosh02, DiVincenzo03, Horodecki03, Ghosh04, Fan04, Horodecki04, Nathanson05, Watrous05, Niset06, Hayashi06, Feng09, Bandyopadhyay10-1, Bandyopadhyay11, Yu12, Yang13, Zhang14, Zhang15, Xu16, Halder18, Halder19, Banik21}. Here the problem arises from the fact that the parts of a composite system are distributed among spatially separated locations and the parties of those locations are allowed to do LOCC only. When it is not possible to identify the state of the system perfectly by LOCC, the given set is called a locally indistinguishable set, otherwise, the set is locally distinguishable set. Clearly, if a given set of orthogonal quantum states is locally indistinguishable then we say that the quantum system exhibit `nonlocality' under the setting of LSDP. This nonlocality is basically the difference between the global and local ability to identify the state of the system. However, apart from examining nonlocal behaviour of a composite quantum system \cite{Bennett99-1, Horodecki03, Bandyopadhyay11, Halder19, Banik21}, the setting of local state discrimination problem is also popular for its applications in data hiding \cite{Terhal01, Eggeling02, Divincenzo02, Lami18, Lami21, Bandyopadhyay21}, and secret sharing \cite{Markham08, Rahaman15, Goswami23}.

For a given locally indistinguishable set, if it is not possible to identify the state of the system under LOCC, even though multiple (finite) identical copies of the states of that set are available, then the given set is locally indistinguishable in many-copy scenario. Note that no adaptive local strategy \cite{Banik21} is used here to examine local indistinguishability in many-copy scenario. However, if the given set contains only orthogonal pure states then the set must be locally distinguishable when sufficient identical copies of the states of the given set are available \cite{Walgate00, Bandyopadhyay11}. Interestingly, it may not be the case when at least one state of the set is a mixed state \cite{Bandyopadhyay11}. This constitutes a fundamental difference between orthogonal pure and mixed states. This difference is described as `more nonlocality with less purity'. Nevertheless, the sets, which are locally indistinguishable in many-copy scenario, are less explored. There are only a few articles \cite{Bandyopadhyay11, Yu14, Li17} where such sets are analyzed. Therefore, there are several questions related to these sets which remained unsolved. In this work we revisit these sets and address several unexplored aspects. In the following, we highlight those aspects demonstrating the main motivations of this work.

Using the properties of unextendible product bases (UPBs), it was shown that there are sets of orthogonal mixed states, which are locally indistinguishable in many-copy scenario \cite{Bandyopadhyay11}. Such a set can be described as the following. Consider a bipartite unextendible product basis (UPB). Then, consider a mixed state which is a normalized projection operator onto the subspace spanned by the product states of the UPB. Along with this mixed state, consider any other state, picked from the entangled subspace, corresponding to the UPB. Then, the set containing these two states is locally indistinguishable in many-copy scenario. After the introduction of such sets, it was an open problem if it is possible to construct such sets without using the notion of UPBs. Later, this question is discussed in \cite{Yu14, Li17}. In these articles, the authors have shown the following without considering the notion of UPBs. Consider any bipartite pure entangled state. Then, in its complementary subspace consider a normalized projection operator such that the pure entangled state and the mixed state in its complementary subspace are supported in the whole Hilbert space. Such an ensemble cannot be perfectly distinguished in many-copy scenario by an operation which is stronger than LOCC. But in these articles the authors have not considered the general notion of UPBs. They have only considered orthogonal UPBs (oUPBs). Clearly, the question remains if there is any connection of the sets of \cite{Yu14, Li17} with nonorthogonal UPBs (nUPBs). Furthermore, to prove that the sets of \cite{Yu14, Li17} are locally indistinguishable in many-copy scenario, the authors have used the notion of unextendibility for operators with positive partial transpose (PPT). Here the question is if one can simplify the proof techniques without using unextendibility for operators with PPT. The main motivation of exploring these questions is to connect several concepts and bringing them together under the setting of LSDP. Furthermore, with better understanding regarding these sets, it may possible to simplify the existing theory. We further ask if it is possible to unify several sets, which are locally indistinguishable in many-copy scenario, by providing a single construction. We mention that given the limited advancement of quantum technologies, unification of several structures is always important as it may provide the opportunity to study several quantum properties using less resources. Then, we ask about constructing sets which are locally indistinguishable in many-copy scenario for multipartite systems and we search for new quantum phenomenon if there is any. We mention that such multipartite sets are not studied before. 

For present type of sets, the notions -- unextendibility, uncompletability play very important roles. These notions were first introduced for product states \cite{Bennett99, DiVincenzo03}. Later, the notion unextendibility has been generalized for a few things, such as, entangled states \cite{Bravyi11}, mutually unbiased bases \cite{Mandayam14}, and operators \cite{Li17}. The notion unextendibility also has a few applications. For example, oUPBs are useful to produce bound entangled states \cite{Bennett99}. Furthermore, oUPBs exhibit nonlocality without entanglement \cite{Bennett99, DiVincenzo03, Rinaldis04, Cohen22}. In Ref.~\cite{DiVincenzo03}, connections of uncompletability and locally distinguishable or indistinguishable sets were explored. Recently, connections between unextendible entangled basis, uncompletable entangled basis and locally indistinguishable sets have been explored in \cite{Halder21, Halder22}. However, there are several questions related to these notions, the answers of which are not known yet. In fact, the notion uncompletability is relatively less explored.

Next, we describe how the rest of this manuscript is arranged and along with this we also discuss our main findings one by one. In Section~\ref{sec2}, we discuss about some definitions along with the settings of LSDP. Then, in the same section we also talk about corresponding assumptions. We want to analyze the role of unextendibility and uncompletability properties to exhibit indistinguishability in many-copy scenario. For this purpose, we first characterize the orthogonal complement of any bipartite pure entangled state and we show that the orthogonal complement of any bipartite pure entangled state is spanned by product states which form a nonorthogonal unextendible product basis (nUPB) of maximum cardinality. This subspace has one to one correspondence with the maximum dimensional subspace where there is no orthonormal product basis. These are discussed in Section~\ref{sec3}. Due to these, for several ensembles, proving local indistinguishability in many-copy scenario becomes simpler. This is explained in Section~\ref{sec4}. Moreover, the following is now clear. There are several ensembles which are locally indistinguishable in many-copy scenario. Construction-wise they are different. But if we consider the technique of proving their local indistinguishability property then, for many of them it can be done using the general notion of UPB. In particular, there are several ensembles of \cite{Yu14, Li17} which have no connection with oUPBs but they are connected with nUPBs. Explicit constructions of the product states, forming nUPBs of maximum cardinality are shown in Section~\ref{sec5}. We also introduce the notion of positive partial transpose uncompletability in Section~\ref{sec6} and examine its role to unify different many-copy indistinguishable ensembles. Then, we report a phenomenon describing more local indistinguishability with less mixed states. This can be found in a class of multipartite ensembles. We mention that this study is a qualitative one because the particular instance that we have considered, there quantitative study is not possible at the moment. The role of bound entanglement in this is also analyzed. These are given in Section~\ref{sec7}. Finally, in Section~\ref{sec8}, the conclusion is drawn.

\section{The task and corresponding assumptions}\label{sec2}
We start with a set of orthogonal quantum states. These states can be pure or mixed states. But only perfect discrimination of these states under LOCC is considered. We also describe the given set as an `ensemble' and within an ensemble, the states are equally probable. The task of distinguishing the states is called local state discrimination problem as we have mentioned it earlier. In other words, it is also called LOCC state discrimination problem. Therefore, it turns out that for a given set of orthogonal states, if it is possible to identify the state of the system perfectly by LOCC, then that set is locally distinguishable set, otherwise, that set is locally indistinguishable. In other words, we say a locally distinguishable set as LOCC distinguishable set and locally indistinguishable set as LOCC indistinguishable set. Corresponding states are called LOCC (or locally) distinguishable and indistinguishable states respectively. When a locally indistinguishable set remains locally indistinguishable even if multiple (finite) identical copies of the states are available, we say that the set is LOCC (or locally) indistinguishable in many-copy scenario. It actually means the following: Suppose, the considered locally indistinguishable set is $\{\rho_1, \rho_2,\dots\}$. In many-copy scenario, $\{\rho_1^{\otimes n}, \rho_2^{\otimes n},\dots\}$ is given, `$n$' is an integer and it is finite. Clearly, the statement `$\{\rho_1, \rho_2,\dots\}$ is locally indistinguishable in many-copy scenario' is equivalent to the statement `$\{\rho_1^{\otimes n}, \rho_2^{\otimes n},\dots\}$ is locally indistinguishable'.

\section{Characterizing subspaces}\label{sec3}
Before we start discussing about the results, we mention that we first talk about bipartite systems. Then, we talk about multipartite systems. We will clearly mention when we will start talking about multipartite systems.

We now ask a couple of questions. However, the first question is related to unextendible product basis (UPB). So, we begin with its definition.

\begin{definition}\label{def1}
[UPB] Suppose, a set of pure states is given. If these states are product and span a proper subspace of the considered Hilbert space such that the complementary subspace contains no product state then these product states form a UPB. 
\end{definition}

If the states within a UPB are pairwise orthogonal to each other then we say that the UPB is an orthogonal UPB (oUPB) \cite{Bennett99, DiVincenzo03}. Otherwise, it is a nonorthogonal UPB (nUPB) \cite{Pittenger03, Parthasarathy04, Bhat06, Leinaas10, Skowronek11}. Note that in this paper if we mention only `UPB' then we actually refer the above definition where the product states may or may not be orthogonal. We assume that the considered system is associated with the Hilbert space, $\mathcal{H}$ = $\mathbb{C}^{d_1}\otimes \mathbb{C}^{d_2}$, $d_1, d_2$ are the dimensions of the local quantum systems. For oUPB, $d_1, d_2\geq3$ \cite{Bennett99, DiVincenzo03}, while for nUPB, there is no such bound. Now, a general question of any bipartite UPB is: 

\begin{question}
What is the maximum cardinality possible for a bipartite UPB?
\end{question}

Here the cardinality or size of a UPB is the number of product states, contained within the UPB. Now if a UPB is an oUPB, then it cannot have cardinality greater than $(d_1d_2-4)$ \cite{Bej21}. This is because in the complementary subspace of an oUPB, there is bound entangled state(s) \cite{Bennett99, Chen11, Halder19-3} which cannot have rank less than four \cite{Horodecki03-1, Chen08}. For the existence of bipartite oUPBs of cardinality $(d_1d_2-4)$, one can have a look into Ref.~\cite{Bej21}. However, for an nUPB, not necessarily the complementary subspace contains a bound entangled state. Therefore, for a bipartite nUPB, the cardinality can be greater than $(d_1d_2-4)$. Here it is important to mention that a bound entangled state is a mixed entangled state from which it is not possible to distill pure state entanglement with some non-zero probability via LOCC, irrespective of how many copies of the state are given \cite{Horodecki97, Horodecki98}.  

We next move to the second question. 

\begin{question}
What is the maximum dimension possible for a bipartite subspace which has no orthonormal product basis?
\end{question}

In general, it is not necessary that a subspace, which has no orthonormal product basis (OPB), must be spanned by the states of an nUPB. For example, there are entangled subspaces which neither have any OPB nor they have any nUPB spanning the subspace. It can also happen that a subspace has an OPB, still the space is spanned by the states of an nUPB. However, when we talk about the existence of maximum cardinality of an nUPB or the existence of the maximum dimension of a subspace which has no OPB, we see that they have one to one correspondence, i.e., Question 1 and Question 2 are connected with each other. In fact, for both questions the answer is $(d_1d_2-1)$ and this connection is through the orthogonal complement of any bipartite pure entangled state. Note that for both questions, the rest of the Hilbert space must be spanned by a pure entangled state which belongs to $\mathcal{H}$ = $\mathbb{C}^{d_1}\otimes \mathbb{C}^{d_2}$, otherwise, the solution might be trivial. More precisely, we want that the one-dimensional subspace, complementary to the nUPB of maximum cardinality, i.e., complementary to the maximum dimensional subspace with no OPB, must be spanned by a pure entangled state with local dimensions $d_1$ or $d_2$. 

We next provide the definition for the orthogonal complement of a bipartite pure state. This definition is required to understand our results. 

\begin{definition}\label{def2}
[Orthogonal complement of a pure state] Given a pure state $\ket{\psi}\in\mathcal{H}$, the orthogonal complement of $\ket{\psi}$ is a proper subspace of $\mathcal{H}$ on which the projection operator $(\mathbf{I}-|\psi\rangle\langle\psi|)$ is supported, $\mathbf{I}$ is the identity operator acting on $\mathcal{H}$.
\end{definition}

We are now ready to present the first proposition:

\begin{proposition}\label{prop1}
The orthogonal complement of any bipartite pure entangled state is spanned by the states of an nUPB of maximum cardinality. In fact, such a subspace, i.e., the orthogonal complement of a pure entangled state, is the only example of maximum dimensional subspace with no OPB. 
\end{proposition}

\noindent
{\bf Proof:} The proof for the first part of the proposition is due to its connection with a separability criterion which states that if the purity of a bipartite mixed state is less than or equal to $1/(d_1d_2-1)$, then it must be a separable state \cite{Gurvits02}, where $d_1d_2$ is the total dimension of the given Hilbert space. We next consider the state $\sigma = [1/(d_1d_2-1)](\mathbf{I}-|\psi\rangle\langle\psi|)$, $\ket{\psi}$ is a pure entangled state. The trace of $\sigma^2$ is given by $1/(d_1d_2-1)$. Therefore, this state is a separable sate, according to the above criterion. Now, the range of a separable state must be spanned by product states \cite{Horodecki97}. This implies that the support of $\sigma$, i.e., the orthogonal complement of any pure entangled state $\ket{\psi}$ is also spanned by a set of product states. These product states must form a UPB because the complementary subspace is an one-dimensional entangled subspace, spanned by $\ket{\psi}$ itself. Clearly, the cardinality of such a UPB is $(d_1d_2-1)$. But oUPB cannot have this cardinality, as described before. Therefore, the product states, spanning the orthogonal complement of any pure entangled state must form an nUPB. Obviously, $(d_1d_2-1)$ is the maximum cardinality of such an nUPB, otherwise, there is no unextendibility. 

The proof for the second part of the proposition is quite straightforward. It is already established that the subspace which is spanned by the states of an nUPB of maximum cardinality, cannot have any OPB, otherwise, they form an oUPB. In this way, nUPB of maximum cardinality provides a sufficient condition for a maximum dimensional subspace with no OPB. Now, notice that if the state $\ket{\psi}$ is a product state then the complementary subspace is trivially spanned by a set of orthogonal product states. Therefore, it is necessary that the maximum dimensional subspace with no OPB must be spanned by the states of an nUPB of maximum cardinality. This completes the proof. $\square$

\begin{remark}\label{rem1}
The pure entangled state in the above proposition is an arbitrary state. Therefore, nUPB of maximum cardinality or the maximum dimensional subspace with no OPB exists in all composite Hilbert spaces. \end{remark}

\begin{remark}\label{rem2}
When the Schmidt rank of the given entangled state $\ket{\psi}$ in the above proposition is strictly greater than two, our subspace coincides with the subspace which does not contain any perfectly LOCC distinguishable basis \cite{Watrous05, Duan09}. However, an important difference is that our subspace exists in every Hilbert space while the subspaces of \cite{Watrous05, Duan09} do not. 
\end{remark}

Here the `Schmidt rank' of a pure entangled state can be defined as the minimum number of product states which are required to express the entangled state. However, in support of Remark \ref{rem2}, one may think about an example. For a two-qubit system, there is no subspace which does not contain any perfectly LOCC distinguishable basis \cite{Watrous05}. But the orthogonal complement of any two-qubit pure entangled state is a subspace with no OPB and it is spanned by the states of an nUPB of cardinality three. 

We now proceed to show that Proposition \ref{prop1} has connection with a class of local state discrimination problems. In fact, due to this proposition, it is possible to provide a simplified theory in those state discrimination problems.

\section{Connection with a class of local state discrimination problems}\label{sec4}
We consider that the state $\ket{\psi}$ is any bipartite pure entangled state. We also consider an arbitrary mixed state $\rho$ of rank $(d_1d_2-1)$ such that it is orthogonal to $\ket{\psi}$, where $d_1d_2$ is the total dimension of the given bipartite Hilbert space. Using existing results, it is possible to prove that the ensemble, $\mathcal{E} \equiv \{|\psi\rangle,~\rho\}$ cannot be perfectly distinguished by LOCC in many-copy scenario. In particular, using the results of \cite{Li17}, it is possible to prove that the states of $\mathcal{E}$ cannot be perfectly distinguished in many-copy scenario by a measurement stronger than LOCC. Such a measurement can be defined through positive operator valued measure (POVM) elements which are positive under partial transpose (PPT). Briefly, we say that the ensemble is indistinguishable under PPT-POVM in many-copy scenario. Now, the measurements, which belong to the LOCC class, are necessarily PPT-POVMs but there are PPT-POVMs which cannot be implemented by LOCC \cite{Yu14}. In this context, we mention that proving (in)distinguishability of the states of any ensemble under LOCC is a difficult task to do. This is a reason why researchers often consider a measurement like PPT-POVM. These measurements have rich mathematical structure and thus, it is relatively easier to prove (in)distinguishability of the states of an ensemble under PPT-POVM. Clearly, if the states of an ensemble are indistinguishable under PPT-POVM then the states must be indistinguishable under LOCC as well. However, before we proceed further, we provide the following in a definition environment. 

\begin{definition}\label{def3}
[PPT-POVM] We consider a measurement on any bipartite Hilbert space. Suppose the measurement is defined by a set of POVM elements such that all of these elements are PPT. Then, we simply say that such a measurement is a PPT-POVM.
\end{definition}

Note that if we consider a measurement according to the above definition, then, this class of measurements is much stronger than LOCC. One may also have a look into \cite{Cheng21} for this class of measurements. Now, we go back to the results of \cite{Li17} again. To prove the local indistinguishability of the states of an ensemble in many-copy scenario, the authors of \cite{Li17} have proved the indistinguishability of the states of the ensemble in many-copy scenario under PPT-POVM first. In fact, for this purpose, they have used the notion of PPT unextendibility. Nevertheless, we argue here that for proving indistinguishability of the states of $\mathcal{E}$ under LOCC in many-copy scenario, it is neither required to use the notion of PPT unextendibility nor one has to demonstrate explicitly that the states of $\mathcal{E}$ cannot be perfectly distinguished by PPT-POVM in many-copy scenario. The proof of local indistinguishability of the states of $\mathcal{E}$ in many-copy scenario can be done with the help of Proposition \ref{prop1}. So, we proceed to prove the following:  

\begin{proposition}\label{prop2}
The states of $\mathcal{E}$ cannot be perfectly distinguished by LOCC in many-copy scenario. 
\end{proposition}

\noindent
{\bf Proof:} The state $\rho$ of $\mathcal{E}$ is supported in a space spanned by the states of an nUPB. This is due to Proposition \ref{prop1}. So, if we consider multiple copies of $\rho$, i.e., $\rho^{\otimes n}$, $n$ is finite, then also this state is supported in the space of nUPB. This is due to the fact that tensor product of two nUPBs is again an nUPB \cite{Cubitt11}. The rest of the proof follows from the arguments given in \cite{Bandyopadhyay11}. These arguments can be summarized as the following. Since the state $\rho^{\otimes n}$ is always supported in a space spanned by the states of some nUPB, the state $|\psi\rangle\langle\psi|^{\otimes n}$ is always supported in an entangled subspace. So, we cannot get a product state which has nonzero overlap with $|\psi\rangle\langle\psi|^{\otimes n}$ and orthogonal to $\rho^{\otimes n}$. Again, finding such a product state is necessary to distinguish the states $\rho^{\otimes n}$ and $|\psi\rangle\langle\psi|^{\otimes n}$ unambiguously under LOCC with at least some nonzero probability \cite{Chefles04, Bandyopadhyay09}. Clearly, unambiguous discrimination of these states is not possible under LOCC with non-zero probability. This implies perfect discrimination of these states is also not possible. In this way, the states of $\mathcal{E}$ cannot be perfectly distinguished by LOCC in many-copy scenario. $\square$ 

Now, regarding the above proof technique, we have the following remark: 

\begin{remark}\label{rem3}
The proof technique for local indistinguishability of $\mathcal{E}$ in many-copy scenario, becomes much simpler and this simplicity is due to the finding, given in Proposition \ref{prop1}.
\end{remark}

Apart from the above simplification, the proof technique of Proposition \ref{prop2} has another importance. This is described as the following. After the introduction of orthogonal quantum states which are locally indistinguishable in many-copy scenario, in Ref.~\cite{Bandyopadhyay11}, it was an open problem how to construct such states without invoking the concept of UPB. This question was addressed later in \cite{Li17}. There, the authors claimed that their ensembles are constructed without using the concept of UPB. However, there the authors only considered the notion of oUPB. Interestingly, our discussions show that several ensembles of \cite{Li17} are connected to nUPB. Therefore, it is still an open problem how to construct such states without invoking the general notion of a UPB. We will come back to this discussion again in a later portion of the paper.  

Notice that Proposition \ref{prop1} only talks about the existence of nUPB of maximum cardinality. In spite of this proof for the existence of an nUPB in the orthogonal complement of any bipartite pure entangled state, the question remains: How to construct such product states starting from an arbitrary bipartite pure entangled state. In the following, we provide a general procedure for constructing the product states, starting from any bipartite pure entangled state. This construction technique is important as it provides the nUPBs of maximum cardinality.

\section{Constructions of nonorthogonal Unextendible Product Bases}\label{sec5}
It is known that any bipartite pure entangled state $\ket{\psi}\in\mathcal{H}=\mathbb{C}^{d_1}\otimes\mathbb{C}^{d_2}$, can be written in the Schmidt form, i.e., $\ket{\psi} = \sum_i a_i\ket{i}\ket{i^\prime}$, where $i=1,\dots,r$, $2\leq r\leq\min\{d_1, d_2\}$, $r$ is known as Schmidt rank, $a_i$ are positive numbers such that $\sum_i a_i^2 = 1$, $a_i$ are called Schmidt coefficients, $\{\ket{i}\}$ are orthonormal vectors for the first subsystem, and $\{\ket{i^\prime}\}$ are orthonormal vectors for the second subsystem. For example, if $r=2$, then $i=1,2$ and the state $\ket{\psi}$ becomes $a_1\ket{1}\ket{1^\prime}+a_2\ket{2}\ket{2^\prime}$. 

Starting from an arbitrary pure entangled state, we show how to construct a set of linearly independent product states which span the orthogonal complement of the given entangled state. Clearly, all these product states must be orthogonal to the given entangled state.

We suppose that the Hilbert space is $\mathcal{H}$ = $\mathbb{C}^2\otimes\mathbb{C}^2$ and $r$ = $2$. So, an arbitrary entangled state is given by- $\ket{\psi}$ = $a_1\ket{1}\ket{1^\prime}+a_2\ket{2}\ket{2^\prime}$. The orthogonal complement of this state is spanned by three states \{$a_2\ket{1}\ket{1^\prime}-a_1\ket{2}\ket{2^\prime}$, $\ket{1}\ket{2^\prime}$, $\ket{2}\ket{1^\prime}$\}. This space is also spanned by three linearly independent states \{$(a_2\ket{1}\ket{1^\prime}+a_2\ket{1}\ket{2^\prime}-a_1\ket{2}\ket{1^\prime}-a_1\ket{2}\ket{2^\prime})/\sqrt{2}$, $\ket{1}\ket{2^\prime}$, $\ket{2}\ket{1^\prime}$\}. In this set, the first product state can be written as $(a_2\ket{1}-a_1\ket{2})(\ket{1^\prime}+\ket{2^\prime})/\sqrt{2}$. It is easy to check that if we pick any two states among these three product states then either the two states are orthogonal to each other or they can be written as $\{\ket{\Phi}, a\ket{\Phi}+a^\prime\ket{\Phi^\prime}\}$, $\langle\Phi|\Phi^\prime\rangle$ = 0. Therefore, the three product states are pairwise linearly independent. Furthermore, all of them are orthogonal to the given entangled state, $\ket{\psi}$. If the Hilbert space dimension is higher but $r$ is still 2, i.e., the state is supported in a smaller dimensional subspace then easily, we can consider some other orthogonal product states along with the above three product states for spanning the whole orthogonal complement of the given entangled state.

Now, we suppose that the Hilbert space is $\mathcal{H}$ = $\mathbb{C}^3\otimes\mathbb{C}^3$ and $r$ = $3$. So, an arbitrary entangled state is given by- $\ket{\psi}$ = $a_1\ket{1}\ket{1^\prime}+a_2\ket{2}\ket{2^\prime}+a_3\ket{3}\ket{3^\prime}$. In this case we consider two subspaces, one is spanned by- \{$(1/N_1)(a_2\ket{1}\ket{1^\prime}-a_1\ket{2}\ket{2^\prime})$, $\ket{1}\ket{2^\prime}$, $\ket{2}\ket{1^\prime}$\} and the other is spanned by- \{$(1/N_2)(a_3\ket{1}\ket{1^\prime}-a_1\ket{3}\ket{3^\prime})$, $\ket{1}\ket{3^\prime}$, $\ket{3}\ket{1^\prime}$\}, $N_1$, $N_2$ are the factors for proper normalization. The first subspace can be spanned by three linearly independent product states, following the $r=2$ case. Again, by the similar procedure, for the second subspace also we can construct three linearly independent product states \{$(1/N_2)(a_3\ket{1}-a_1\ket{3})(\ket{1^\prime}+\ket{3^\prime})/\sqrt{2}$, $\ket{1}\ket{3^\prime}$, $\ket{3}\ket{1^\prime}$\}. So, we now have a set of six linearly independent product states, together they are given by- \{$(1/N_1)(a_2\ket{1}-a_1\ket{2})(\ket{1^\prime}+\ket{2^\prime})/\sqrt{2}$, $\ket{1}\ket{2^\prime}$, $\ket{2}\ket{1^\prime}$, $(1/N_2)(a_3\ket{1}-a_1\ket{3})(\ket{1^\prime}+\ket{3^\prime})/\sqrt{2}$, $\ket{1}\ket{3^\prime}$, $\ket{3}\ket{1^\prime}$\}. But for the entangled state that we have considered, the orthogonal complement is an eight dimensional subspace. So, we need two more product states which are given by- \{$\ket{2}\ket{3^\prime}$, $\ket{3}\ket{2^\prime}$\}. In this way we can construct linearly independent product states which span the orthogonal complement of the state $\ket{\psi}$ = $a_1\ket{1}\ket{1^\prime}+a_2\ket{2}\ket{2^\prime}+a_3\ket{3}\ket{3^\prime}$. If the Hilbert space dimension is higher but $r$ is still 3, i.e., the state is supported in a smaller dimensional subspace then quite easily we can consider some other orthogonal product states along with the above eight product states for spanning the whole orthogonal complement of the given entangled state.

Following similar process, one can easily construct such product states which span the orthogonal complement of a bipartite entangled state with an arbitrary value of $r$. For any $\mathcal{H}$ = $\mathbb{C}^{d_1}\otimes\mathbb{C}^{d_2}$ and an arbitrary $r(\geq3)$, the key is to consider separate three-dimensional subspaces spanned by the states \{$(1/N_{(i-1)})(a_i\ket{1}\ket{1^\prime}-a_1\ket{i}\ket{i^\prime})$, $\ket{1}\ket{i^\prime}$, $\ket{i}\ket{1^\prime}$\} which is also spanned by the product states \{$(1/N_{(i-1)})(a_i\ket{1}-a_1\ket{i})(\ket{1^\prime}+\ket{i^\prime})/\sqrt{2}$, $\ket{1}\ket{i^\prime}$, $\ket{i}\ket{1^\prime}$\}, for different values of $i=2,3,\dots,r$, $N_{(i-1)}$ are appropriate factors for proper normalizations. For a fixed value of $r$, we can construct $3(r-1)$ such product states and with these we can add some orthogonal product states (total $r^2-3r+2$ product states) $\ket{i_1}\ket{i_2^\prime}$, $i_1\neq i_2\neq1$, $i_1$, $i_2$ = $2,3,\dots,r$. To complete the set, the rest of the product states are given by- \{$\ket{j_1}\ket{j_2^\prime}$\} $\setminus$ \{$\ket{i_1}\ket{i_2^\prime}$\}, where $j_1$ = $1,2,\dots d_1$, $j_2$ = $1,2,\dots d_2$, and $i_1$, $i_2$ = $1,2,\dots,r$.

These constructions are also important because following these constructions, it is possible to construct nUPBs, having cardinality $<(d_1d_2-1)$. We now provide an interesting example of that kind. 

We consider $\mathcal{H} = \mathbb{C}^2\otimes\mathbb{C}^3$. Then, we consider the construction of the subspace which contain only state with non-positive partial transpose (NPT) \cite{Johnston13}. Briefly, we say that such a subspace is an NPT subspace. We consider the span of two entangled states, $\ket{1}\ket{2}-\ket{2}\ket{1}$ and $\ket{1}\ket{3}-\ket{2}\ket{2}$. For simplicity, we avoid the factors for normalization. Now, this span of two entangled states produces an NPT subspace while the complementary subspace is spanned by the states \{$\ket{1}\ket{2}+\ket{2}\ket{1}$, $\ket{1}\ket{3}+\ket{2}\ket{2}$, $\ket{1}\ket{1}$, $\ket{2}\ket{3}$\}. We take combinations of these four states and construct linearly independent product states. These product states are \{$(\ket{1}+\ket{2})(\ket{1}+\ket{2}+\ket{3})$, $(\ket{1}-\ket{2})(\ket{1}-\ket{2}+\ket{3})$, $\ket{1}\ket{1}$, $\ket{2}\ket{3}$\} (without the factors for normalization). It is easy to check that these product states are linearly independent and these product states are orthogonal to the entangled states of the considered NPT subspace. In this way, we find an nUPB of cardinality $(d_1d_2-2)$, where $d_1=2$ and $d_2=3$. This construction is particularly important because there is no oUPB in $\mathcal{H} = \mathbb{C}^2\otimes\mathbb{C}^d$, $d\geq2$ \cite{DiVincenzo03}. 

We next consider any mixed state of rank-4 which is supported on the subspace spanned by the states of nUPB, just constructed. Then, this mixed state along with any orthogonal state(s) from the NPT subspace form an ensemble, the states of which cannot be perfectly distinguished by LOCC in many-copy scenario. The proof of this follows the same way as given for Proposition \ref{prop2}.

Using oUPBs, it is possible to construct ensembles which cannot be perfectly distinguished by LOCC in many-copy scenario \cite{Bandyopadhyay11}. On the other hand, one can consider the ensembles of \cite{Li17, Yu14}, they are also locally indistinguishable in many-copy scenario. Interestingly, many of them are connected with nUPBs. This is due to the proof of Proposition \ref{prop2}. We have already discussed about this before starting this section. Thus, we have the following remark: 

\begin{remark}\label{rem4}
There are several ensembles which are locally indistinguishable in many-copy scenario. Construction-wise they are different. But if we consider the technique of proving their local indistinguishability property then, for many of them it can be done using the general notion of UPB.
\end{remark}
 
However, the ensembles of \cite{Bandyopadhyay11} are perfectly distinguishable by PPT-POVM, one can have a look into \cite{Cheng21}. On the other hand, if we consider the ensembles of \cite{Yu14, Li17}, the states are indistinguishable under PPT-POVM. Therefore, we ask the following. We want to construct a single UPB (say, UPB with {\it common property} or cUPB), starting from which we further want to construct several ensembles, the states of which cannot be perfectly distinguished by LOCC in many-copy scenario. But it must be the case that for some ensembles the states are perfectly PPT-POVM distinguishable while for other ensembles, it is not. For this purpose, we introduce the concept `PPT uncompletability'.

\section{PPT uncompletability}\label{sec6}
PPT uncompletability is a property which can be exhibited by a PPT uncompletable subspace. PPT uncompletability is a natural generalization of uncompletability of product states \cite{DiVincenzo03}. Consider a set of orthogonal product states. These product states span a proper subspace of the considered Hilbert space. Now we suppose that there are product state(s) orthogonal to the given product states but it is not possible to extend the set of given states to a complete orthogonal product basis. Then, the given product states exhibit uncompletability. Such product states have connection with local indistinguishability property. For a detailed discussion, one can have a look into \cite{DiVincenzo03}. Next, we start with the following definition for operators with positive partial transpose, i.e., PPT operator.

\begin{definition}\label{def4}
[PPT uncompletable subspace] Let $\mathcal{S}$ be a proper subspace of the bipartite Hilbert space $\mathcal{H}$ such that it contains at least one PPT operator, supported on the entire subspace. Again, we consider that the complementary subspace $\mathcal{S}^\perp$ has PPT operator(s) but none of these operators is supported on the entire $\mathcal{S}^\perp$. If this is the case then we say that the subspace $\mathcal{S}$ is a PPT uncompletable subspace.
\end{definition}

If the PPT operator(s) of $\mathcal{S}^\perp$ are supported in $\mathcal{S}^\prime$, a proper subspace of $\mathcal{S}^\perp$, then $\mathcal{S}\oplus\mathcal{S}^\prime$ is a PPT unextendible subspace \cite{Li17}. Recall that the type of UPB, we just talked about, for that UPB (in particular, cUPB), PPT uncompletability is a desired property. This is because PPT completability may lead to perfect discrimination by PPT POVM. Because of PPT completeness, it may possible to find PPT operators, summing which one may get identity operator. Thus, these operators may constitute a valid PPT POVM which can distinguish some ensembles. On the other hand, PPT unextendibility leads to indistinguishability by PPT POVM. Clearly, PPT uncompletability lies in between PPT completability and PPT unextendibility. We now provide a construction of such a subspace and the cUPB. In fact, this construction is an application of nUPB of maximum cardinality.

\begin{figure}[t!]
\centering
\includegraphics[width = 0.6\textwidth]{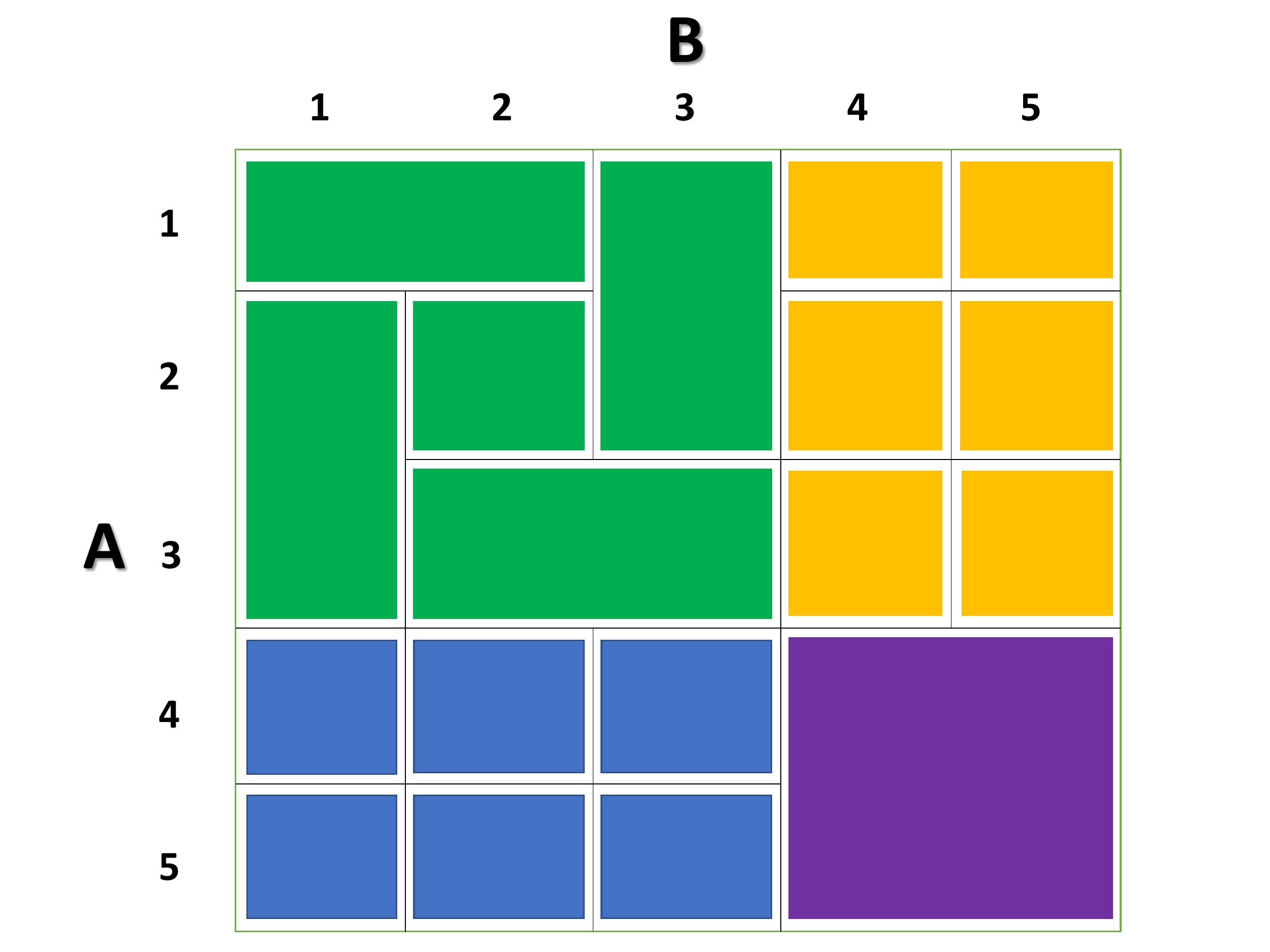}
\fcaption{\label{fig1} Tile structure for $\mathbf{S}$: the green, blue, orange, and the purple colors represent $S_1$, $S_2$, $S_3$, and $S_4$ respectively. In $S_1$ there is a tile structure corresponding to the well-known Tiles UPB \cite{Bennett99}. In $S_2$ and in $S_3$ each tile corresponds to a product state. But in $S_4$ the product states are nonorthogonal. So, there is no tile in this region. Note that `A' stands for Alice's side and `B' stands for Bob's side.}
\end{figure}

We consider $\mathcal{H}$ = $\mathbb{C}^5\otimes\mathbb{C}^5$. We further consider four subspaces $S_1$, $S_2$, $S_3$, and $S_4$, such that $S_1\oplus S_2\oplus S_3\oplus S_4$ = $\mathbb{C}^5\otimes\mathbb{C}^5$. We now define the subspaces in the following way: $S_1$ = $\{\ket{1},\ket{2},\ket{3}\}\otimes\{\ket{1},\ket{2},\ket{3}\}$, $S_2$ = $\{\ket{4},\ket{5}\}\otimes\{\ket{1},\ket{2},\ket{3}\}$, $S_3$ = $\{\ket{1},\ket{2},\ket{3}\}\otimes\{\ket{4},\ket{5}\}$, and $S_4$ = $\{\ket{4},\ket{5}\}\otimes\{\ket{4},\ket{5}\}$, where $S_i$ = $\{\ket{j}\}\otimes\{\ket{k}\}$ means the subspace $S_i$ is spanned by the states $\{\ket{j}\ket{k}\}$. We next consider an oUPB in $S_1$. The states of this oUPB is given by- \{$\ket{1}(\ket{1}-\ket{2})$, $(\ket{1}-\ket{2})\ket{3}$, $\ket{3}(\ket{2}-\ket{3})$, $(\ket{2}-\ket{3})\ket{1}$, $(\ket{1}+\ket{2}+\ket{3})(\ket{1}+\ket{2}+\ket{3})$\}. This oUPB can be found in \cite{Bennett99}. For simplicity, we avoid the factors for normalization here. Again, we consider an nUPB of maximum cardinality in $S_4$, defined by- \{$(\ket{4}+\ket{5})(\ket{4}+\ket{5})$, $\ket{4}\ket{5}$, $\ket{5}\ket{4}$\}. We now consider the following product states together: the product states of oUPB of $S_1$, the product states \{$\ket{4}\ket{1}$, $\ket{4}\ket{2}$, $\ket{4}\ket{3}$, $\ket{5}\ket{1}$, $\ket{5}\ket{2}$, $\ket{5}\ket{3}$\} forming an OPB for $S_2$, the product states \{$\ket{1}\ket{4}$, $\ket{2}\ket{4}$, $\ket{3}\ket{4}$, $\ket{1}\ket{5}$, $\ket{2}\ket{5}$, $\ket{3}\ket{5}$\} forming an OPB for $S_3$, and the product states forming the nUPB of $S_4$ and we say these product states as the set $\mathbf{S}$ (see Fig.~\ref{fig1}). We next prove the following: 

\begin{proposition}\label{prop3}
The product states of $\mathbf{S}$ form an nUPB of cardinality twenty in $\mathbb{C}^5\otimes\mathbb{C}^5$, i.e., the complementary subspace of this nUPB is a five-dimensional entangled subspace. Furthermore, this nUPB constitutes an example of PPT uncompletable subspace.
\end{proposition}

\noindent
{\bf Proof:} The five-dimensional complementary subspace is formed due to the four-dimensional entangled subspace, complementary to the oUPB of $S_1$ and one dimensional entangled subspace, complementary to the nUPB of $S_4$. This one dimensional entangled subspace of $S_4$ is spanned by $\ket{4}\ket{4}-\ket{5}\ket{5}$. Any state which belong to any of these entangled subspaces must be entangled. So, we consider only those states which are supported in both entangled subspace. For these states, Alice or Bob can get pure state entanglement with some nonzero probability performing a simple projective measurement: \{$P_1$ = $|1\rangle\langle1|+|2\rangle\langle2|+|3\rangle\langle3|$, $P_2$ = $|4\rangle\langle4|+|5\rangle\langle5|$; $P_1 + P_2$ = $\mathbf{I}_5$\}, where $\mathbf{I}_5$ is an identity operator acting on $\mathbb{C}^5$. In particular, if the measurement outcome is `2', then a state will be projected on $\ket{4}\ket{4}-\ket{5}\ket{5}$. Thus, all such states are entangled and the five-dimensional complementary subspace is an entangled subspace.

Following the above arguments, it is possible to explain the construction of a PPT uncompletable subspace. The five dimensional entangled subspace, mentioned above, does not have any rank-5 operator which is PPT. Because from all such operators, it is possible to get pure state entanglement, notice the consequence of outcome `2' in the above paragraph. Now, there is a rank-4 PPT operator in this subspace which is due to the oUPB of $S_1$ \cite{Bennett99}. Next, using the states of nUPB of size twenty, it is possible to construct a PPT state supported on the entire nUPB space by taking any convex combination of the product states of nUPB. Therefore, this nUPB space has at least one PPT operator supported on the entire subspace. On the other hand, the complementary subspace contains a PPT operator but there is no PPT operator which is supported on the entire complementary subspace. So, by Definition \ref{def4}, this nUPB space is an example of PPT uncompletable subspace. $\square$

The above nUPB is the desired cUPB (UPB with {\it common property}). Thus, we also prove the following:

\begin{proposition}\label{prop4}
The product states of $\mathbf{S}$ lead to a single UPB, from which both types of many-copy indistinguishable ensembles can be constructed; one is perfectly distinguishable by PPT-POVM while the other cannot be. 
\end{proposition}

\noindent
{\bf Note:} In this proposition, when we consider (in)distinguishability of these ensembles under PPT-POVM, we examine this at a single copy level.

\noindent
{\bf Proof:} We consider any state $\rho_1$ of rank-20 supported on the nUPB space of size twenty. Then, we consider $\rho_2$ which is the rank four PPT entangled state, supported in the complementary subspace of the above nUPB. We also consider $\rho_3$ which is the projector on the state $\ket{4}\ket{4}-\ket{5}\ket{5}$. Following the proof of Proposition \ref{prop2}, it is easy to see that the states of both ensembles $\{\rho_1, \rho_2\}$, $\{\rho_1, \rho_3\}$ cannot be perfectly distinguished by LOCC in many-copy scenario. 

But the states of the ensemble $\{\rho_1, \rho_2\}$ can be perfectly distinguished by PPT-POVM. Corresponding measurement is given by $\{P^\prime, \mathbf{I}_{25}-P^\prime\}$, where $P^\prime$ is the projection operator onto the support of $\rho_2$ and $\mathbf{I}_{25}$ is the identity operator acting on $\mathbb{C}^5\otimes\mathbb{C}^5$ (see also \cite{Cheng21} in this regard). Notice that the space on which the operator $\mathbf{I}_{25}-P^\prime$ is supported, is spanned by the states of an oUPB. This oUPB can be found by considering the oUPB of $S_1$ together with any OPBs of $S_2$, $S_3$, and $S_4$. Clearly, $\mathbf{I}_{25}-P^\prime$ must be a PPT operator. Again, $P^\prime$ is the projection operator corresponding to the bound entangled state which is produced due to the oUPB of $S_1$. Thus, $P^\prime$ is also a PPT operator. In this way, $\{P^\prime, \mathbf{I}_{25}-P^\prime\}$ constitutes a PPT-POVM. 

Furthermore, the states of the ensemble $\{\rho_1, \rho_3\}$ cannot be perfectly distinguished by PPT-POVM. Because for this perfect discrimination, we need a PPT operator orthogonal to $\rho_1$ but nonorthogonal to $\rho_3$. Such an operator must be partially supported on the space spanned by $\ket{4}\ket{4}-\ket{5}\ket{5}$. However, we have proved that such an operator does not exist. $\square$ 

As described before, both $\{\rho_1,~\rho_2\}$, $\{\rho_1,~\rho_3\}$ cannot be perfectly distinguished by LOCC in many-copy scenario but they have different properties when PPT-POVM is concerned. Thus, we can have the following remark.

\begin{remark}\label{rem5}
The notion of PPT uncompletability is helping to unify different indistinguishability property through a single construction of a UPB, in particular, the cUPB. 
\end{remark}

\section{More local indistinguishability with less mixed states}\label{sec7}
Here we discuss about the results for multipartite quantum systems. Given an ensemble of multipartite orthogonal quantum states, for local indistinguishability of the ensemble, it is not necessary that the ensemble must be locally indistinguishable across at least one bipartition. For example, one can consider the following: any set of orthogonal product states in three-qubit Hilbert space, is perfectly locally distinguishable in bipartitions, however, such a set might be locally indistinguishable when all qubits are spatially separated \cite{Bennett99, DiVincenzo03, Halder19, Halder20}. On the other hand, if any multipartite ensemble is locally indistinguishable across at least one bipartition then it must be so when all subsystems are spatially separated. Therefore, one can say the following: 

\begin{remark}\label{rem6}
Local indistinguishability across bipartitions is a stronger property compared to local indistinguishability when all subsystems are spatially separated. 
\end{remark}

We now want to talk about a class of many-copy indistinguishable multipartite ensembles which are defined as the following: there are only two elements in those ensembles, i.e., $\{\sigma_1, \sigma_2\}$, such that they are orthogonal to each other and the sum of the dimensions of the supports, corresponding to the states $\sigma_1$ and $\sigma_2$, is equal to the total dimension of the considered Hilbert space. In fact, here $\sigma_1$ is a separable state while $\sigma_2$ is an entangled state. For these ensembles, we present the following proposition:

\begin{proposition}\label{prop5}
If $\sigma_2$ is a pure state, then, the ensemble must be locally indistinguishable in many-copy scenario across at least one bipartition. 
\end{proposition}

\noindent
{\bf Proof:} A pure multipartite entangled state is entangled in at least one bipartition. We start with that bipartition and one of the states in the ensemble becomes a pure bipartite entangled state while the other state of the ensemble is supported in the entire orthogonal complement of that pure state. The rest of the proof is due to Proposition \ref{prop1} and Proposition \ref{prop2}. $\square$ 

So, in the above, we see that pure state entanglement guarantees local indistinguishability in many-copy scenario in at least one bipartition for the present class of states. Now we prove the following: 

\begin{proposition}\label{prop6}
If $\sigma_2$ is a mixed state, then, the ensemble may not be locally indistinguishable in many-copy scenario across bipartitions.
\end{proposition}

\noindent
{\bf Proof:} To prove this proposition, any example, where the ensemble is perfectly locally distinguishable across bipartitions but it is locally indistinguishable in many-copy scenario when all subsystems are spatially separated, is sufficient, provided $\sigma_2$ is a mixed state. Such examples are given as the following. 

We consider any three-qubit oUPB. We then consider any mixed separable state supported in the entire subspace spanned by the states of the oUPB. We say this state as $\sigma_1$. Then, we consider another state, the bound entangled state, corresponding to the oUPB. We say this state as $\sigma_2$. Now, the ensemble $\{\sigma_1, \sigma_2\}$ is locally indistinguishable in many-copy scenario when all qubits are spatially separated. The proof corresponding to this follows from the construction of \cite{Bandyopadhyay11}. Nevertheless, it is important to mention that though the results of \cite{Bandyopadhyay11} are for bipartite systems, they can be easily extended to multipartite systems, following the fact that if we take tensor product of multipartite oUPBs, then after the tensor product, another multipartite oUPB is produced \cite{DiVincenzo03}. So, here what we are considering is the following. $\sigma_1$ is supported on the entire oUPB space. Now, if we take $\sigma_1^{\otimes n}$, then it is also supported on some oUPB space, due to \cite{DiVincenzo03}. Therefore, $\sigma_2^{\otimes n}$ is always supported in an entangled subspace. Thus, the ensemble $\{\sigma_1^{\otimes n}, \sigma_2^{\otimes n}\}$ is locally indistinguishable \cite{Bandyopadhyay11}. We next proceed to prove that the ensemble $\{\sigma_1, \sigma_2\}$ are perfectly distinguishable across bipartitions. 

It is known that the bound entangled state, $\sigma_2$ is separable across every bipartition \cite{Bennett99}. Since, we are starting with any three-qubit oUPB, it is difficult to give an explicit form. But the above can be understood as the following. If we consider the oUPB, due to which the bound entangled entangled state is produced, then it can extended to a complete orthogonal product basis across bipartitions. Therefore, this bound entangled state can be written as a convex combination of orthogonal product states across every bipartition. However, these product states are not fully product states. Thereafter, in any bipartition, we consider the orthogonal product states together, which span the supports of the mixed states $\sigma_1$ and $\sigma_2$. These pure states can be chosen in such a way that they form a complete OPB (cOPB) in $\mathbb{C}^2\otimes\mathbb{C}^4$. If one changes the bipartition, the cOPB also changes. These cOPBs are perfectly distinguishable by LOCC \cite{Bennett99}. In this way, the ensemble $\{\sigma_1, \sigma_2\}$ is perfectly locally distinguishable across every bipartition. These complete the proof of the proposition. $\square$ 

These sets are strange sets: when all qubits are spatially separated, the ensemble is quite indistinguishable under LOCC as they remain indistinguishable in many-copy scenario. Nonetheless, when considered in the bipartitions with only one copy of each state is given, the set is perfectly locally distinguishable, i.e., the local indistinguishability property completely vanishes. Therefore, the local indistinguishability property here is fragile with respect to bipartitioning. This observation is completely new and therefore, the above proposition is quite important.

Recall that any set of three-qubit orthogonal product states is perfectly locally distinguishable across bipartitions. However, the situation becomes complex, when there are mixed states. But for distinguishability of the present ensembles across bipartitions, the bound entangled state is playing a key role as it is separable across every bipartition. We like to analyze this to a further extend in a later portion of this paper. 

In light of Proposition \ref{prop5} and Proposition \ref{prop6}, when we analyze these ensembles, we see that the local indistinguishability property of them may increase with decreasing number of mixed states in those ensembles. This is in the following sense. A stronger form of local indistinguishability (see Remark \ref{rem6}) can be guaranteed when a state of the present ensembles is a pure state, i.e., local indistinguishability across at least one bipartition in many-copy scenario can be guaranteed. But this property may not be found when both states are mixed states. 

This is not usual as `more mixed states' in an ensemble usually provides `less distinguishability' of the ensemble under LOCC. There are reasons to believe it. For example, any two orthogonal pure states can always be perfectly distinguished by LOCC \cite{Walgate00}. However, there are ensembles of two orthogonal mixed states which cannot be perfectly distinguished by LOCC \cite{Halder21}. Again, any set of orthogonal pure states can always be perfectly distinguished by LOCC if sufficient (finite) identical copies of the states are available \cite{Walgate00, Bandyopadhyay11}. But there are orthogonal mixed states which cannot be perfectly distinguished by LOCC in many-copy scenario \cite{Bandyopadhyay11, Duan09, Li17}. From these results, one may expect that if the number of mixed states increases in an ensemble then the local indistinguishability property of the set is also increased. However, in the present class of ensembles, we see that a stronger demonstration of local indistinguishability can only be guaranteed if one of the states within the ensemble is a pure state. On the other hand, this notion may not be found if both states are mixed states. Clearly, with decreasing number of mixed states in the present class of ensembles, one may get more local indistinguishability property. We say this phenomenon as {\it more local indistinguishability with less mixed states}. 

We mention that our analysis is a qualitative analysis. In fact, this analysis is complementary to the previous qualitative result `more nonlocality with less purity' \cite{Bandyopadhyay11}. We also mention that the property we are discussing about here, i.e., the stronger demonstration of local indistinguishability property within the present class of ensembles, can only be guaranteed when one of the states in present ensembles is a pure state. In fact, this particular demonstration of local indistinguishability property has nothing to do with the quantitative measure of mixedness within the states of the ensemble but with the quality of a mixed state. To explore this further, we analyze the role of bound entanglement in exhibiting `more local indistinguishability with less mixed states'. This can be realized through the following remark. 

\begin{remark}\label{rem7}
Bound entanglement via positivity under partial transpose may not be desired to exhibit more local indistinguishability in these ensembles.
\end{remark}

We start by providing a proposition.

\begin{proposition}\label{prop7}
To distinguish any ensemble $\{\sigma_1, \sigma_2\}$ of the present kind across all bipartitions by LOCC, it is necessary that the projection operators corresponding to $\sigma_1$ and $\sigma_2$ are positive under partial transpose in bipartitions. 
\end{proposition}

\noindent
{\bf Note:} When we say `ensemble of present kind', we actually refer to the description given just before Proposition \ref{prop5}.

\noindent
{\bf Proof:} To distinguish any ensemble $\{\sigma_1, \sigma_2\}$ perfectly by LOCC, it is necessary that the ensemble is perfectly distinguishable by PPT-POVM. For this, we need two projection operators $\pi_1$ and $\pi_2$, such that Tr$(\sigma_i\pi_j)$ = $\delta_{ij}$; $\pi_1+\pi_2$ = $\mathbf{I}$, here $\mathbf{I}$ is the identity operator, acting on the Hilbert space corresponding to $\sigma_1$ and $\sigma_2$. We note that $\pi_i$ must be PPT in bipartitions here $\forall i$ = $1,2$. 

We now assume that a projector is non-positive under partial transpose (NPT). For this, it is necessary that there exist at least one bipartition in which the projector is NPT. In this way, in that bipartition the ensemble $\{\sigma_1, \sigma_2\}$ is not perfectly distinguishable by PPT-POVM and thereby LOCC. $\square$ 

From the above one thing is clear: to distinguish present sets in bipartitions, it is necessary that $\sigma_2$ has bound entanglement via positive partial transpose particularly when it is a normalized projection operator. Therefore, one can conclude from here that bound entanglement may not be desired to exhibit {\it more local indistinguishability} in those cases. This is in the following sense. If the normalized projector corresponding to $\sigma_2$ has bound entanglement then there is a possibility that the ensemble is perfectly locally distinguishable across every bipartition, in this context, recall the examples of three-qubit system, discussed earlier.

\section{Conclusion and open problems}\label{sec8}
Apart from exploring the notions -- unextendibility for product states and uncompletability for PPT operators, here we have discussed several topics which were not addressed before in the literature. These topics cover connection of nonorthogonal UPB with several many-copy indistinguishable ensembles, unifying several many-copy indistinguishable ensembles, role of PPT uncompletability in this unification, exhibiting new phenomenon via many-copy indistinguishable ensembles in multipartite systems and role of bound entanglement in it. In particular, we have shown that the orthogonal complement of any bipartite pure entangled state is spanned by product states which form an nUPB of maximum cardinality. This subspace has one to one correspondence with the maximum dimensional subspace where there is no OPB. Due to this, for several ensembles, proving local indistinguishability in many-copy scenario has become simpler. Explicit construction of the product states, forming the nUPBs of maximum cardinality, has been shown. In fact, the following is now clear. There are several many-copy indistinguishable ensembles under LOCC. Construction-wise they are different. But if we consider the technique of proving their local indistinguishability property then, for many of them it can be done using the general notion of UPB. Then, we have introduced the notion of PPT uncompletability and via this notion we have unified different many-copy indistinguishable ensembles. Finally, we have discussed about a class of many-copy indistinguishable multipartite ensembles for which local indistinguishability property increases with decreasing number of mixed states in the ensemble.

For further research, it will be interesting to identify the ensembles which are not connected to the general notion of UPB, but still the states of such an ensemble are locally indistinguishable in many-copy scenario. It will also be interesting to explore more about the new notion PPT uncompletability and the phenomenon increasing local indistinguishability with decreasing number of mixed states in an ensemble.

\nonumsection{Acknowledgements}
\noindent
This work was supported by the National Science Centre, Poland, within the QuantERA II Programme (No2021/03/Y/ST2/00178, acronym ExTRaQT) that has received funding from the European Union's Horizon 2020 research and innovation programme under Grant Agreement No 101017733 and the ``Quantum Optical Technologies'' project, carried out within the International Research Agendas programme of the Foundation for Polish Science co-financed by the European Union under the European Regional Development Fund.

\nonumsection{References}
\bibliographystyle{rinton_press}
\bibliography{ref}

\begin{thebibliography}{10}
\providecommand{\url}[1]{\texttt{#1}}
\providecommand{\urlprefix}{URL }
\expandafter\ifx\csname urlstyle\endcsname\relax
  \providecommand{\doi}[1]{doi:\discretionary{}{}{}#1}\else
  \providecommand{\doi}{doi:\discretionary{}{}{}\begingroup
  \urlstyle{rm}\Url}\fi

\bibitem{Bennett99-1}
C.~H. Bennett, D.~P. DiVincenzo, C.~A. Fuchs, T.~Mor, E.~Rains, P.~W. Shor,
  J.~A. Smolin and W.~K. Wootters (1999), \emph{Quantum nonlocality without
  entanglement}, Phys. Rev. A, vol.~59, pp. 1070--1091,
  \doi{10.1103/PhysRevA.59.1070},
  \urlprefix\url{https://link.aps.org/doi/10.1103/PhysRevA.59.1070}.

\bibitem{Bennett99}
C.~H. Bennett, D.~P. DiVincenzo, T.~Mor, P.~W. Shor, J.~A. Smolin and B.~M.
  Terhal (1999), \emph{Unextendible Product Bases and Bound Entanglement},
  Phys. Rev. Lett., vol.~82, pp. 5385--5388, \doi{10.1103/PhysRevLett.82.5385}.

\bibitem{Walgate00}
J.~Walgate, A.~J. Short, L.~Hardy and V.~Vedral (2000), \emph{Local
  Distinguishability of Multipartite Orthogonal Quantum States}, Phys. Rev.
  Lett., vol.~85, pp. 4972--4975, \doi{10.1103/PhysRevLett.85.4972},
  \urlprefix\url{https://link.aps.org/doi/10.1103/PhysRevLett.85.4972}.

\bibitem{Virmani01}
S.~Virmani, M.~F. Sacchi, M.~B. Plenio and D.~Markham (2001), \emph{Optimal
  local discrimination of two multipartite pure states}, Phys. Lett. A., vol.
  288, p.~62, \doi{https://doi.org/10.1016/S0375-9601(01)00484-4}.

\bibitem{Ghosh01}
S.~Ghosh, G.~Kar, A.~Roy, A.~Sen(De) and U.~Sen (2001),
  \emph{Distinguishability of Bell States}, Phys. Rev. Lett., vol.~87, p.
  277902, \doi{10.1103/PhysRevLett.87.277902},
  \urlprefix\url{https://link.aps.org/doi/10.1103/PhysRevLett.87.277902}.

\bibitem{Walgate02}
J.~Walgate and L.~Hardy (2002), \emph{Nonlocality, Asymmetry, and
  Distinguishing Bipartite States}, Phys. Rev. Lett., vol.~89, p. 147901,
  \doi{10.1103/PhysRevLett.89.147901},
  \urlprefix\url{https://link.aps.org/doi/10.1103/PhysRevLett.89.147901}.

\bibitem{Ghosh02}
S.~Ghosh, G.~Kar, A.~Roy, D.~Sarkar, A.~Sen(De) and U.~Sen (2002), \emph{Local
  indistinguishability of orthogonal pure states by using a bound on
  distillable entanglement}, Phys. Rev. A, vol.~65, p. 062307,
  \doi{10.1103/PhysRevA.65.062307},
  \urlprefix\url{https://link.aps.org/doi/10.1103/PhysRevA.65.062307}.

\bibitem{DiVincenzo03}
D.~P. DiVincenzo, T.~Mor, P.~W. Shor, J.~A. Smolin and B.~M. Terhal (2003),
  \emph{Unextendible Product Bases, Uncompletable Product Bases and Bound
  Entanglement}, Commun. Math. Phys., vol. 238, pp. 379--410,
  \doi{10.1007/s00220-003-0877-6}.

\bibitem{Horodecki03}
M.~Horodecki, A.~Sen(De), U.~Sen and K.~Horodecki (2003), \emph{Local
  Indistinguishability: More Nonlocality with Less Entanglement}, Phys. Rev.
  Lett., vol.~90, p. 047902, \doi{10.1103/PhysRevLett.90.047902},
  \urlprefix\url{https://link.aps.org/doi/10.1103/PhysRevLett.90.047902}.

\bibitem{Ghosh04}
S.~Ghosh, G.~Kar, A.~Roy and D.~Sarkar (2004), \emph{Distinguishability of
  maximally entangled states}, Phys. Rev. A, vol.~70, p. 022304,
  \doi{10.1103/PhysRevA.70.022304},
  \urlprefix\url{https://link.aps.org/doi/10.1103/PhysRevA.70.022304}.

\bibitem{Fan04}
H.~Fan (2004), \emph{Distinguishability and Indistinguishability by Local
  Operations and Classical Communication}, Phys. Rev. Lett., vol.~92, p.
  177905, \doi{10.1103/PhysRevLett.92.177905},
  \urlprefix\url{https://link.aps.org/doi/10.1103/PhysRevLett.92.177905}.

\bibitem{Horodecki04}
M.~Horodecki, J.~Oppenheim, A.~Sen(De) and U.~Sen (2004), \emph{Distillation
  Protocols: Output Entanglement and Local Mutual Information}, Phys. Rev.
  Lett., vol.~93, p. 170503, \doi{10.1103/PhysRevLett.93.170503},
  \urlprefix\url{https://link.aps.org/doi/10.1103/PhysRevLett.93.170503}.

\bibitem{Nathanson05}
M.~Nathanson (2005), \emph{Distinguishing bipartite orthogonal states by LOCC:
  best and worst cases}, J. Math. Phys., vol.~46, p. 062103,
  \doi{https://doi.org/10.1063/1.1914731}.

\bibitem{Watrous05}
J.~Watrous (2005), \emph{Bipartite Subspaces Having No Bases Distinguishable by
  Local Operations and Classical Communication}, Phys. Rev. Lett., vol.~95, p.
  080505, \doi{10.1103/PhysRevLett.95.080505},
  \urlprefix\url{https://link.aps.org/doi/10.1103/PhysRevLett.95.080505}.

\bibitem{Niset06}
J.~Niset and N.~J. Cerf (2006), \emph{Multipartite nonlocality without
  entanglement in many dimensions}, Phys. Rev. A, vol.~74, p. 052103,
  \doi{10.1103/PhysRevA.74.052103},
  \urlprefix\url{https://link.aps.org/doi/10.1103/PhysRevA.74.052103}.

\bibitem{Hayashi06}
M.~Hayashi, D.~Markham, M.~Murao, M.~Owari and S.~Virmani (2006), \emph{Bounds
  on Multipartite Entangled Orthogonal State Discrimination Using Local
  Operations and Classical Communication}, Phys. Rev. Lett., vol.~96, p.
  040501, \doi{10.1103/PhysRevLett.96.040501},
  \urlprefix\url{https://link.aps.org/doi/10.1103/PhysRevLett.96.040501}.

\bibitem{Feng09}
Y.~{Feng} and Y.~{Shi} (2009), \emph{Characterizing Locally Indistinguishable
  Orthogonal Product States}, IEEE Trans. Inf. Theory, vol.~55, pp. 2799--2806,
  ISSN 0018-9448, \doi{10.1109/TIT.2009.2018330}.

\bibitem{Bandyopadhyay10-1}
S.~Bandyopadhyay (2010), \emph{Entanglement and perfect discrimination of a
  class of multiqubit states by local operations and classical communication},
  Phys. Rev. A, vol.~81, p. 022327, \doi{10.1103/PhysRevA.81.022327},
  \urlprefix\url{https://link.aps.org/doi/10.1103/PhysRevA.81.022327}.

\bibitem{Bandyopadhyay11}
S.~Bandyopadhyay (2011), \emph{More Nonlocality with Less Purity}, Phys. Rev.
  Lett., vol. 106, p. 210402, \doi{10.1103/PhysRevLett.106.210402},
  \urlprefix\url{https://link.aps.org/doi/10.1103/PhysRevLett.106.210402}.

\bibitem{Yu12}
N.~Yu, R.~Duan and M.~Ying (2012), \emph{Four Locally Indistinguishable
  Ququad-Ququad Orthogonal Maximally Entangled States}, Phys. Rev. Lett., vol.
  109, p. 020506, \doi{10.1103/PhysRevLett.109.020506},
  \urlprefix\url{https://link.aps.org/doi/10.1103/PhysRevLett.109.020506}.

\bibitem{Yang13}
Y.-H. Yang, F.~Gao, G.-J. Tian, T.-Q. Cao and Q.-Y. Wen (2013), \emph{Local
  distinguishability of orthogonal quantum states in a
  $2\ensuremath{\bigotimes}2\ensuremath{\bigotimes}2$ system}, Phys. Rev. A,
  vol.~88, p. 024301, \doi{10.1103/PhysRevA.88.024301},
  \urlprefix\url{https://link.aps.org/doi/10.1103/PhysRevA.88.024301}.

\bibitem{Zhang14}
Z.-C. Zhang, F.~Gao, G.-J. Tian, T.-Q. Cao and Q.-Y. Wen (2014),
  \emph{Nonlocality of orthogonal product basis quantum states}, Phys. Rev. A,
  vol.~90, p. 022313, \doi{10.1103/PhysRevA.90.022313},
  \urlprefix\url{https://link.aps.org/doi/10.1103/PhysRevA.90.022313}.

\bibitem{Zhang15}
Z.-C. Zhang, F.~Gao, S.-J. Qin, Y.-H. Yang and Q.-Y. Wen (2015),
  \emph{Nonlocality of orthogonal product states}, Phys. Rev. A, vol.~92, p.
  012332, \doi{10.1103/PhysRevA.92.012332},
  \urlprefix\url{https://link.aps.org/doi/10.1103/PhysRevA.92.012332}.

\bibitem{Xu16}
G.-B. Xu, Q.-Y. Wen, S.-J. Qin, Y.-H. Yang and F.~Gao (2016), \emph{Quantum
  nonlocality of multipartite orthogonal product states}, Phys. Rev. A,
  vol.~93, p. 032341, \doi{10.1103/PhysRevA.93.032341},
  \urlprefix\url{https://link.aps.org/doi/10.1103/PhysRevA.93.032341}.

\bibitem{Halder18}
S.~Halder (2018), \emph{Several nonlocal sets of multipartite pure orthogonal
  product states}, Phys. Rev. A, vol.~98, p. 022303,
  \doi{10.1103/PhysRevA.98.022303},
  \urlprefix\url{https://link.aps.org/doi/10.1103/PhysRevA.98.022303}.

\bibitem{Halder19}
S.~Halder, M.~Banik, S.~Agrawal and S.~Bandyopadhyay (2019), \emph{Strong
  Quantum Nonlocality without Entanglement}, Phys. Rev. Lett., vol. 122, p.
  040403, \doi{10.1103/PhysRevLett.122.040403},
  \urlprefix\url{https://link.aps.org/doi/10.1103/PhysRevLett.122.040403}.

\bibitem{Banik21}
M.~Banik, T.~Guha, M.~Alimuddin, G.~Kar, S.~Halder and S.~S. Bhattacharya
  (2021), \emph{Multicopy Adaptive Local Discrimination: Strongest Possible
  Two-Qubit Nonlocal Bases}, Phys. Rev. Lett., vol. 126, p. 210505,
  \doi{10.1103/PhysRevLett.126.210505},
  \urlprefix\url{https://link.aps.org/doi/10.1103/PhysRevLett.126.210505}.

\bibitem{Terhal01}
B.~M. Terhal, D.~P. DiVincenzo and D.~W. Leung (2001), \emph{Hiding Bits in
  Bell States}, Phys. Rev. Lett., vol.~86, pp. 5807--5810,
  \doi{10.1103/PhysRevLett.86.5807},
  \urlprefix\url{https://link.aps.org/doi/10.1103/PhysRevLett.86.5807}.

\bibitem{Eggeling02}
T.~Eggeling and R.~F. Werner (2002), \emph{Hiding Classical Data in
  Multipartite Quantum States}, Phys. Rev. Lett., vol.~89, p. 097905,
  \doi{10.1103/PhysRevLett.89.097905},
  \urlprefix\url{https://link.aps.org/doi/10.1103/PhysRevLett.89.097905}.

\bibitem{Divincenzo02}
D.~P. {DiVincenzo}, D.~W. {Leung} and B.~M. {Terhal} (2002), \emph{Quantum data
  hiding}, IEEE Trans. Inf. Theory, vol.~48, pp. 580--598,
  \doi{10.1109/18.985948}.

\bibitem{Lami18}
L.~Lami, C.~Palazuelos and A.~Winter (2018), \emph{Ultimate Data Hiding in
  Quantum Mechanics and Beyond}, Commun. Math. Phys., vol. 361, pp. 661 -- 708,
  \doi{10.1007/s00220-018-3154-4},
  \urlprefix\url{https://doi.org/10.1007/s00220-018-3154-4}.

\bibitem{Lami21}
L.~Lami (2021), \emph{Quantum data hiding with continuous-variable systems},
  Phys. Rev. A, vol. 104, p. 052428, \doi{10.1103/PhysRevA.104.052428},
  \urlprefix\url{https://link.aps.org/doi/10.1103/PhysRevA.104.052428}.

\bibitem{Bandyopadhyay21}
S.~Bandyopadhyay and S.~Halder (2021), \emph{Genuine activation of nonlocality:
  From locally available to locally hidden information}, Phys. Rev. A, vol.
  104, p. L050201, \doi{10.1103/PhysRevA.104.L050201},
  \urlprefix\url{https://link.aps.org/doi/10.1103/PhysRevA.104.L050201}.

\bibitem{Markham08}
D.~Markham and B.~C. Sanders (2008), \emph{Graph states for quantum secret
  sharing}, Phys. Rev. A, vol.~78, p. 042309, \doi{10.1103/PhysRevA.78.042309},
  \urlprefix\url{https://link.aps.org/doi/10.1103/PhysRevA.78.042309}.

\bibitem{Rahaman15}
R.~Rahaman and M.~G. Parker (2015), \emph{Quantum scheme for secret sharing
  based on local distinguishability}, Phys. Rev. A, vol.~91, p. 022330,
  \doi{10.1103/PhysRevA.91.022330},
  \urlprefix\url{https://link.aps.org/doi/10.1103/PhysRevA.91.022330}.

\bibitem{Goswami23}
S.~Goswami and S.~Halder (2023), \emph{Information locking and its
  resource-efficient extraction}, Phys. Rev. A, vol. 108, p. 012405,
  \doi{10.1103/PhysRevA.108.012405},
  \urlprefix\url{https://link.aps.org/doi/10.1103/PhysRevA.108.012405}.

\bibitem{Yu14}
N.~{Yu}, R.~{Duan} and M.~{Ying} (2014), \emph{Distinguishability of Quantum
  States by Positive Operator-Valued Measures with Positive Partial Transpose},
  IEEE Trans. Inf. Theory, vol.~60, pp. 2069--2079, ISSN 0018-9448,
  \doi{10.1109/TIT.2014.2307575}.

\bibitem{Li17}
Y.~Li, X.~Wang and R.~Duan (2017), \emph{Indistinguishability of bipartite
  states by positive-partial-transpose operations in the many-copy scenario},
  Phys. Rev. A, vol.~95, p. 052346, \doi{10.1103/PhysRevA.95.052346},
  \urlprefix\url{https://link.aps.org/doi/10.1103/PhysRevA.95.052346}.

\bibitem{Bravyi11}
S.~Bravyi and J.~A. Smolin (2011), \emph{Unextendible maximally entangled
  bases}, Phys. Rev. A, vol.~84, p. 042306, \doi{10.1103/PhysRevA.84.042306},
  \urlprefix\url{https://link.aps.org/doi/10.1103/PhysRevA.84.042306}.

\bibitem{Mandayam14}
P.~Mandayam, S.~Bandyopadhyay, M.~Grassl and W.~K. Wootters (2014),
  \emph{Unextendible mutually unbiased bases from Pauli Classes}, Quantum Inf.
  Comput., vol.~14, p. 823, \doi{10.26421/QIC14.9-10-8}.

\bibitem{Rinaldis04}
S.~De~Rinaldis (2004), \emph{Distinguishability of complete and unextendible
  product bases}, Phys. Rev. A, vol.~70, p. 022309,
  \doi{10.1103/PhysRevA.70.022309},
  \urlprefix\url{https://link.aps.org/doi/10.1103/PhysRevA.70.022309}.

\bibitem{Cohen22}
S.~M. Cohen (2022), \emph{Local approximation of multipartite quantum
  measurements}, Phys. Rev. A, vol. 105, p. 022207,
  \doi{10.1103/PhysRevA.105.022207},
  \urlprefix\url{https://link.aps.org/doi/10.1103/PhysRevA.105.022207}.

\bibitem{Halder21}
S.~Halder and U.~Sen (2021), \emph{Local indistinguishability and
  incompleteness of entangled orthogonal bases: Method to generate two-element
  locally indistinguishable ensembles}, Ann. Phys., vol. 431, p. 168550, ISSN
  0003-4916, \doi{https://doi.org/10.1016/j.aop.2021.168550},
  \urlprefix\url{https://www.sciencedirect.com/science/article/pii/S0003491621001561}.

\bibitem{Halder22}
S.~Halder and U.~Sen (2022), \emph{Unextendible entangled bases and more
  nonlocality with less entanglement}, Phys. Rev. A, vol. 105, p. L030401,
  \doi{10.1103/PhysRevA.105.L030401},
  \urlprefix\url{https://link.aps.org/doi/10.1103/PhysRevA.105.L030401}.

\bibitem{Pittenger03}
A.~O. Pittenger (2003), \emph{Unextendible product bases and the construction
  of inseparable states}, Linear Alg. Appl., vol. 359, pp. 235--248, ISSN
  0024-3795, \doi{https://doi.org/10.1016/S0024-3795(02)00423-8},
  \urlprefix\url{https://www.sciencedirect.com/science/article/pii/S0024379502004238}.

\bibitem{Parthasarathy04}
K.~R. Parthasarathy (2004), \emph{On the maximal dimension of a completely
  entangled subspace for finite level quantum systems}, Proc. Math. Sci., vol.
  114, pp. 365--374, \doi{10.1007/BF02829441}.

\bibitem{Bhat06}
B.~V.~R. Bhat (2006), \emph{A COMPLETELY ENTANGLED SUBSPACE OF MAXIMAL
  DIMENSION}, Int. J. Quantum Inform., vol.~04, pp. 325--330,
  \doi{10.1142/S0219749906001797}.

\bibitem{Leinaas10}
J.~M. Leinaas, J.~Myrheim and P.~O. Sollid (2010), \emph{Low-rank extremal
  positive-partial-transpose states and unextendible product bases}, Phys. Rev.
  A, vol.~81, p. 062330, \doi{10.1103/PhysRevA.81.062330},
  \urlprefix\url{https://link.aps.org/doi/10.1103/PhysRevA.81.062330}.

\bibitem{Skowronek11}
{\L}.~Skowronek (2011), \emph{Three-by-three bound entanglement with general
  unextendible product bases}, J. Math. Phys., vol.~52, p. 122202,
  \doi{10.1063/1.3663836}, \urlprefix\url{https://doi.org/10.1063/1.3663836}.

\bibitem{Bej21}
P.~Bej and S.~Halder (2021), \emph{Unextendible product bases, bound entangled
  states, and the range criterion}, Phys. Lett. A, vol. 386, p. 126992, ISSN
  0375-9601, \doi{https://doi.org/10.1016/j.physleta.2020.126992},
  \urlprefix\url{https://www.sciencedirect.com/science/article/pii/S0375960120308598}.

\bibitem{Chen11}
L.~Chen and D.~Z. Djokovic (2011), \emph{Description of rank four entangled
  states of two qutrits having positive partial transpose}, J. Math. Phys.,
  vol.~52, p. 122203, \doi{10.1063/1.3663837},
  \urlprefix\url{https://doi.org/10.1063/1.3663837}.

\bibitem{Halder19-3}
S.~Halder, M.~Banik and S.~Ghosh (2019), \emph{Family of bound entangled states
  on the boundary of the Peres set}, Phys. Rev. A, vol.~99, p. 062329,
  \doi{10.1103/PhysRevA.99.062329},
  \urlprefix\url{https://link.aps.org/doi/10.1103/PhysRevA.99.062329}.

\bibitem{Horodecki03-1}
P.~Horodecki, J.~A. Smolin, B.~M. Terhal and A.~V. Thapliyal (2003), \emph{Rank
  two bipartite bound entangled states do not exist}, Theor. Comput. Sci., vol.
  292, pp. 589--596, \doi{10.1016/S0304-3975(01)00376-0}.

\bibitem{Chen08}
L.~Chen and Y.-X. Chen (2008), \emph{Rank-three bipartite entangled states are
  distillable}, Phys. Rev. A, vol.~78, p. 022318,
  \doi{10.1103/PhysRevA.78.022318}.

\bibitem{Horodecki97}
P.~Horodecki (1997), \emph{Separability criterion and inseparable mixed states
  with positive partial transposition}, Phys. Lett. A, vol. 232, pp. 333 --
  339, ISSN 0375-9601, \doi{https://doi.org/10.1016/S0375-9601(97)00416-7},
  \urlprefix\url{http://www.sciencedirect.com/science/article/pii/S0375960197004167}.

\bibitem{Horodecki98}
M.~Horodecki, P.~Horodecki and R.~Horodecki (1998), \emph{Mixed-State
  Entanglement and Distillation: Is there a ``Bound'' Entanglement in Nature?},
  Phys. Rev. Lett., vol.~80, pp. 5239--5242, \doi{10.1103/PhysRevLett.80.5239}.

\bibitem{Gurvits02}
L.~Gurvits and H.~Barnum (2002), \emph{Largest separable balls around the
  maximally mixed bipartite quantum state}, Phys. Rev. A, vol.~66, p. 062311,
  \doi{10.1103/PhysRevA.66.062311},
  \urlprefix\url{https://link.aps.org/doi/10.1103/PhysRevA.66.062311}.

\bibitem{Duan09}
R.~Duan, Y.~Feng, Y.~Xin and M.~Ying (2009), \emph{Distinguishability of
  Quantum States by Separable Operations}, IEEE Trans. Inf. Theory, vol.~55,
  pp. 1320--1330, \doi{10.1109/TIT.2008.2011524}.

\bibitem{Cheng21}
H.-C. Cheng, A.~Winter and N.~Yu (2021), \emph{Discrimination of quantum states
  under locality constraints in the many-copy setting}, in \emph{2021 IEEE
  International Symposium on Information Theory (ISIT)}, pp. 1188--1193,
  \doi{10.1109/ISIT45174.2021.9518100}.

\bibitem{Cubitt11}
T.~S. Cubitt, J.~Chen and A.~W. Harrow (2011), \emph{Superactivation of the
  Asymptotic Zero-Error Classical Capacity of a Quantum Channel}, IEEE Trans.
  Inf. Theory, vol.~57, pp. 8114--8126, \doi{10.1109/TIT.2011.2169109}.

\bibitem{Chefles04}
A.~Chefles (2004), \emph{Condition for unambiguous state discrimination using
  local operations and classical communication}, Phys. Rev. A, vol.~69, p.
  050307(R), \doi{10.1103/PhysRevA.69.050307},
  \urlprefix\url{https://link.aps.org/doi/10.1103/PhysRevA.69.050307}.

\bibitem{Bandyopadhyay09}
S.~Bandyopadhyay and J.~Walgate (2009), \emph{Local distinguishability of any
  three quantum states}, J. Phys. A: Math. Theor., vol.~42, p. 072002,
  \doi{http://iopscience.iop.org/1751-8121/42/7/072002}.

\bibitem{Johnston13}
N.~Johnston (2013), \emph{Non-positive-partial-transpose subspaces can be as
  large as any entangled subspace}, Phys. Rev. A, vol.~87, p. 064302,
  \doi{10.1103/PhysRevA.87.064302},
  \urlprefix\url{https://link.aps.org/doi/10.1103/PhysRevA.87.064302}.

\bibitem{Halder20}
S.~Halder and R.~Sengupta (2020), \emph{Distinguishability classes, resource
  sharing, and bound entanglement distribution}, Phys. Rev. A, vol. 101, p.
  012311, \doi{10.1103/PhysRevA.101.012311},
  \urlprefix\url{https://link.aps.org/doi/10.1103/PhysRevA.101.012311}.

\end{thebibliography}
\end{document}